\documentstyle[12pt,graphicx]{article}

\newcommand{\publication}{JETP Letters {\bf 80}, 343 (2004)}           
\title{\vspace*{-2.5cm}{\Large\bf
Quantum phase transition for the BEC--BCS\\ 
crossover in condensed matter physics and\\ 
CPT violation in elementary particle physics}}
\author{
F.R. Klinkhamer\,\thanks{Email address: frans.klinkhamer@physik.uni-karlsruhe.de}\\
{\normalsize 
Institute for Theoretical Physics, University of Karlsruhe (TH),}\\
{\normalsize 76128 Karlsruhe, Germany}\\ 
and\\
G.E. Volovik\,\thanks{Email address: volovik@boojum.hut.fi}\\
{\normalsize Low Temperature Laboratory, Helsinki University of Technology,}\\
{\normalsize P.O. Box 2200, FIN--02015 HUT, Finland;}\\
{\normalsize L.D. Landau Institute for Theoretical Physics, 
Russian Academy of Sciences,}\\
{\normalsize Kosygina 2, 117940 Moscow, Russia}}

\date{\vspace*{0.25cm}{\normalsize
[\publication; cond-mat/0407597]}}

\begin{document}
\onecolumn

\maketitle
\begin{abstract}
{\noindent We discuss the quantum phase transition that separates 
a vacuum state with fully-gapped fermion spectrum  from a vacuum state
with topologically-protected Fermi points (gap nodes). In the
context of condensed-matter physics, such a quantum phase
transition with Fermi point splitting may occur for a system of
ultracold fermionic atoms in the region of the BEC--BCS
crossover, provided Cooper pairing occurs in the non-$s$-wave
channel. For elementary particle physics, the splitting of Fermi
points may lead to CPT violation, neutrino
oscillations, and other phenomena.}
\end{abstract}

\def\g{\kappa}
\def\half{{1\over2}}
\def\L{{\mathcal L}}
\def\S{{\mathcal S}}
\def\d{{\mathrm{d}}}
\def\x{{\mathbf x}}
\def\v{{\mathbf v}}
\def\im{{\rm i}}
\def\etal{{\emph{et al\/}}}
\def\det{{\mathrm{det}}}
\def\tr{{\mathrm{tr}}}
\def\ie{{\emph{i.e.}}}
\def\bnabla{\mbox{\boldmath$\nabla$}}
\def\Box{\kern0.5pt{\lower0.1pt\vbox{\hrule height.5pt width 6.8pt
     \hbox{\vrule width.5pt height6pt \kern6pt \vrule width.3pt}
     \hrule height.3pt width 6.8pt} }\kern1.5pt}
\def\HRULE{{\bigskip\hrule\bigskip}}
\def\be{\begin{equation}}
\def\ee{\end{equation}}
\def\implies{\Rightarrow}
\newcommand {\CSlike}{Chern--Simons-like}
\newcommand {\CS}    {Chern--Simons}
\newcommand {\PV}    {Pauli--Villars}
\newcommand {\LC}    {Levi--Civita}
\newcommand {\BN}    {Bogoliubov--Nambu}
\newcommand {\WZNW}  {Wess--Zumino--Novikov--Witten}
\newcommand {\BCS}   {Bardeen--Cooper--Schrieffer}
\newcommand {\cgth}  {chiral gauge theory}
\newcommand {\qfth}  {quantum field theory}
\newcommand {\qfths} {quantum field theories}
\newcommand {\rqfth} {relativistic quantum field theory}
\newcommand {\rqfths}{relativistic quantum field theories}

\newpage
There are two major schemes for the classification of states in
condensed matter physics and relativistic quantum field theory:
classification by symmetry and by universality classes. 

For the first classification method, a given state of the
system is characterized by a symmetry group $H$ which is a subgroup
of the symmetry group $G$ of the relevant physical laws (see,
e.g., Ref. \cite{VolovikGorkov1985} for symmetry classification of
superconducting states). The thermodynamic phase transition
between equilibrium states is usually marked by a change of the
symmetry group $H$. The subgroup $H$ is also responsible for 
topological defects,  which are determined by the nontrivial
elements of the homotopy groups $\pi_n(G/H)$; cf. Ref.
\cite{TopologyReview1}.

The second classification method deals with the ground states of the
system at zero temperature ($T=0$), i.e., it is the classification
of quantum vacua. The universality class  determines the general
features of the quantum vacuum,  such as the
linear response and the energy
spectrum of fermionic excitations.  For translation-invariant
systems in which momentum is a well-defined quantity,  these
features of the fermionic quantum vacuum are determined by
momentum-space topology. For (3+1)--dimensional systems, there
are only three basic universality classes of fermionic vacua
\cite{VolovikBook}: (i) vacua with  fully-gapped fermionic
excitations; (ii) vacua with fermionic excitations characterized
by Fermi points (the excitations behave  as massless Weyl fermions close
to the Fermi points); (iii)  vacua with fermionic excitations
characterized by Fermi surfaces.
[Fermi points ${\bf p}_n$
are points in 3-momentum space at which the energy vanishes,
$E({\bf p}_n)=0$, and similarly for Fermi surfaces $S_n$, with
$E({\bf p})=0$ for  ${\bf p} \in S_n$.] 

It may happen that by changing some parameter $q$ of the system we
transfer the vacuum state from one universality class  to
another, without changing its symmetry group $H$. The point
$q_c$, where this zero-temperature transition occurs, marks the quantum
phase transition. For $T\neq 0$, the phase transition is absent,
as the two states belong to the same symmetry class $H$. Hence,
there is an isolated singular point $(q_c,0)$ in the $(q,T)$
plane. Two examples of a quantum phase transition are (i) the
Lifshitz transition in crystals, at which the Fermi surface changes
its topology or shrinks to a point, and (ii) the transition between
states with different values of the Hall (or spin-Hall)
conductance in (2+1)--dimensional systems.

\begin{figure}[p]
\centerline{\includegraphics[width=0.85\linewidth]{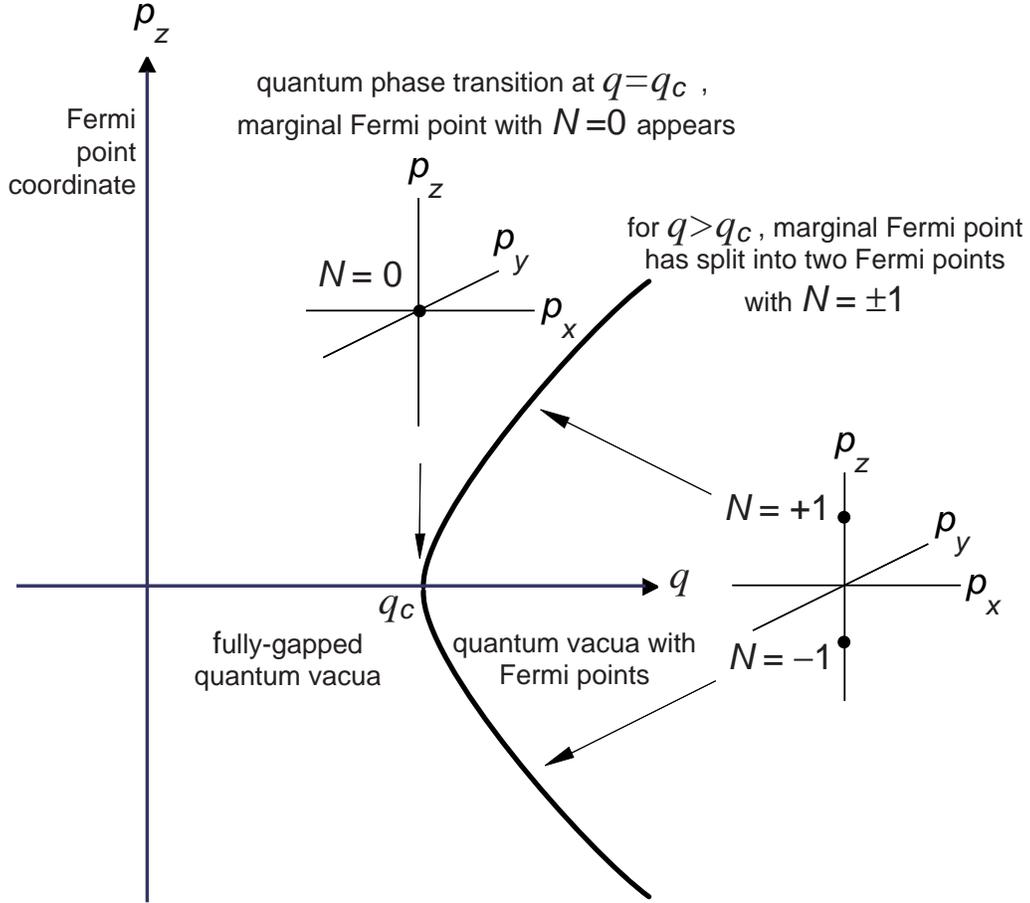}} 
\vspace*{3\baselineskip} 
\caption{Quantum phase transition at
$q=q_c$ between a fully-gapped vacuum and a vacuum with
topologically-protected Fermi points (gap nodes). At $q=q_c$, 
there appears a marginal Fermi point  with topological charge $N=0$ (inset at
the top). 
For $q > q_c$, the marginal Fermi point has split into two Fermi points
characterized by nonzero topological invariants $N = \pm 1$
(inset on the right). For a system  of ultracold fermionic atoms
qualitatively  described by Hamiltonians (\ref{BogoliubovNambuH})
and (\ref{BogoliubovNambuAlphaH}), the critical parameter is
$q_c=0$ [note that eight Fermi points emerge  
for the case of Hamiltonian (\ref{BogoliubovNambuAlphaH})]. 
For Dirac fermions with CPT violation in
Hamiltonian (\ref{ModifiedHDirac}), the parameter $q$ is chosen as
$q\equiv |{\bf b}|$ and the critical parameter is $q_c=M$.}
         \label{Fig1}
\end{figure}

In this Letter, we discuss the quantum phase transition between
a vacuum with fully-gapped fermionic excitations  and a vacuum
with Fermi points. At the transition point $q=q_c$, a
topologically-trivial Fermi point emerges from the fully-gapped
state. This marginal Fermi point then splits into two or more
topologically-nontrivial Fermi points (see Fig.~\ref{Fig1}).
The topologically-protected Fermi points give rise to anomalous
properties of the system in the low-temperature regime;
cf. Sec.~7.3.2 of  Ref.~\cite{VollhardtWoelfle} and
Part IV of Ref.~\cite{VolovikBook}.

These effects may  occur in a system of ultracold fermionic atoms
in the region of the BEC--BCS crossover in a non-$s$--wave Cooper channel. 
Superfluidity in the BEC regime and the BEC--BCS
crossover has been observed for $^{40}$K and $^6$Li atoms
\cite{FAP1,FAP2,FAP3,FAP4,FAP5}.  In these experiments, a
magnetic-field Feshbach resonance was used to control the
interactions in the $s$-wave channel. For the case of $s$-wave
pairing, there are fully-gapped vacua on both sides of the
crossover and there is no quantum phase transition. If, however, the
pairing occurs in a non-$s$-wave channel, a quantum phase
transition may be expected between the fully-gapped state and the
state with Fermi points. It was reported recently
\cite{PWaveResonance1,PWaveResonance2} that three $p$--wave Feshbach 
resonances were found for $^6$Li atoms. This suggests the possibility of
future observations of non-$s$-wave pairing and of the quantum phase
transition associated with the splitting of Fermi points.

Here, we will discuss two examples of such a transition, using for
simplicity $p$--wave spin-triplet pairing and their possible
analogs in relativistic quantum field theory. 
We also argue in the following 
that a similar quantum phase transition characterized by Fermi point
splitting may occur for the Standard Model of elementary particle 
physics \cite{EWSM}, but refer the reader to 
Refs.~\cite{SplittingPreprint,KlinkhamerJETPL,KlinkhamerNeutrino}
for further details.  In fact, condensed-matter
physics provides us with a broad class of \qfths~not
restricted by Lorentz invariance, which allows us to consider many
problems in the \rqfth~of the Standard Model 
from a more general perspective. Just as
for nonrelativistic systems, the basic properties of relativistic
\qfths~(including quantum anomalies) are determined by 
momentum-space topology, which classifies the relativistic vacua
according to the same three universality classes.

Since we are only interested in effects determined by the topology
and the symmetry of the fermionic Green's function $G(p)$, we do not 
require a special
form of the Green's function and can choose the simplest one
with the required topology. First, consider the
\BN~Hamiltonian which qualitatively describes fermionic
quasiparticles in the axial state of $p$--wave pairing.
This Hamiltonian can be applied  to both the Bardeen--Cooper--Schrieffer
(BCS) and Bose--Einstein condensation (BEC) regimes, and also to
superfluid $^3$He-A \cite{VollhardtWoelfle}. Specifically, the 
\BN~Hamiltonian is given by:
\begin{eqnarray}   
\!\!\!\!\!\!\!\!&&\!\!\!\!\!\!\!\!
H=  \nonumber\\[4mm]
\!\!\!\!\!\!\!\!&&\!\!\!\!\!\!\!\!
\left(\matrix{ |{\bf p}|^2/ 2m  -q
&c_\perp\,{\bf p}\cdot (\widehat{\bf e}_1+ i\, \widehat{\bf e}_2) \cr
    c_\perp\,{\bf p}\cdot (\widehat{\bf e}_1- i\, \widehat{\bf e}_2)
&-|{\bf p}|^2/ 2m +q \cr }\right),
\label{BogoliubovNambuH}
\end{eqnarray}
and $G^{-1}(i\omega,{\bf p})=i\omega - H({\bf p})$, with $\hbar=1$.
Considered are fermionic atoms of mass $m$
with a given direction of the
atomic spin, assuming that only these atoms experience the
Feshbach resonance. 
The orthonormal triad $(\widehat{\bf e}_1,\, \widehat{\bf e}_2,\,
\widehat{\bf l}\equiv \widehat{\bf e}_1\times \widehat{\bf e}_2)$ and
the maximum transverse speed $c_\perp$ of the quasiparticles
characterize the order parameter in the axial state of triplet
superfluid. The unit vector $\widehat{\bf l}$ corresponds to the
direction of the orbital momentum of the Cooper pair or the
diatomic molecule. 
We further assume that the parameter $q$ is controlled by the magnetic
field in the vicinity of the Feshbach resonance. 

The energy spectrum of these  \BN~fermions is
\begin{equation}
E^2 ({\bf p}) = \left({|{\bf p}|^2\over 2m}-q\right)^2+\,
          c_\perp^2\,\left({\bf p}\times \widehat{\bf l}\,\right)^2 .
\label{BogoliubovNambuE}
\end{equation}
The BCS regime occurs for $q>0$, with the parameter $q$ playing the 
role of a chemical potential. In this regime, there are two Fermi points,
i.e., points in 3-momentum space with $E({\bf p})=0$. For the energy
spectrum (\ref{BogoliubovNambuE}), the Fermi points are 
${\bf p}_1= p_F \, \widehat{\bf l}$ 
and ${\bf p}_2=-p_F \, \widehat{\bf l}$, with Fermi momentum
$p_F=\sqrt{2 m q}$.

For a general system, be it relativistic or nonrelativistic, the
stability of the $a$-th Fermi point is guaranteed by the
topological invariant $N_{a}$, which can be written as a surface
integral in frequency-momentum space. In terms of the fermionic
propagator $G=G(p_0,p_1,p_2,p_3)$, for $p_\mu=(\omega,{\bf p})$, the
topological invariant is \cite{VolovikBook}
\begin{eqnarray}    
\!\!\!\!\!\!&&\!\!\!\!\!\!
N_{a} \equiv {1\over{24\pi^2}}\,\epsilon_{\mu\nu\rho\sigma}~
{\rm tr} \,\oint_{\Sigma_a} dS^{\sigma}  
\nonumber\\[2mm]
\!\!\!\!\!\!&&\!\!\!\!\!\!
\times\,
G\frac{\partial}{\partial p_\mu} G^{-1}\;
G\frac{\partial}{\partial p_\nu} G^{-1}\;
G\frac{\partial}{\partial p_\rho}G^{-1},
\label{TopInvariant}
\end{eqnarray}
where $\Sigma_a$ is a three-dimensional surface around the
isolated Fermi point $p_{\mu a}=(0,{\bf p}_a)$ and `tr' stands for 
the trace over the relevant spin indices.

For the case considered, the trace in Eq.~(\ref{TopInvariant})
is over the Bogoliubov-Nambu spin and the
two Fermi points  ${\bf p}_1$ and  ${\bf p}_2$
have nonzero topological charges $N_1=+1$ and
$N_2=-1$. The density of states in this gapless regime is given by 
$\nu(E)\propto E^2$. At $q=0$, these two Fermi points merge and
form one topologically-trivial Fermi point with $N=0$. This
intermediate state, 
which appears at the quantum phase transition 
($q_c =0$), is marginal: the momentum-space
topology is trivial and cannot protect the vacuum against
decay into one of the two topologically-stable vacua. For $q<0$,
the marginal Fermi point disappears altogether and the spectrum
becomes fully-gapped. In this topologically-stable fully-gapped
vacuum, the density of states is drastically different from that
in the topologically-stable gapless regime: $\nu(E)=0$ for
$E<|q|$. All this demonstrates that the quantum phase transition
considered is of purely topological origin.

Note that if a single pair of Fermi points appears in momentum
space, the vacuum state has nonzero internal angular momentum
along ${\hat{\bf l}}$, i.e., this quantum vacuum has the property of an
orbital ferromagnet. Later, we will discuss an example with
multiple Fermi points, for which the total orbital momentum is
zero and the vacuum state corresponds to an orbital
antiferromagnet.

We now turn to elementary particle physics \cite{EWSM}.
It appears that the vacu\-um of the Stan\-dard Mo\-del 
above the electroweak transition (vanishing fermion masses) is
marginal:  there is a multiply degenerate Fermi point ${\bf
p}=0$ with topological charge $N=0$. It is therefore the
intermediate state between two topologically-stable vacua, the
fully-gapped vacuum and the vacuum with topologically-nontrivial
Fermi points.  In the Standard Model, this marginal Fermi point is
protected by symmetries, namely the continuous electroweak
symmetry (or the discrete symmetry discussed in Sec. 12.3.2 of
Ref.\cite{VolovikBook}) and the CPT symmetry. 

Explicit violation 
or spontaneous breaking of one of these symmetries transforms the
marginal vacuum of the Standard Model
into one of the two topologically-stable vacua.  If, 
for example, the electroweak symmetry is broken, the marginal Fermi point
disappears and the fermions become massive. This is known to happen 
in the case of  
quarks and electrically charged leptons below the
electroweak transition.  If, on the other hand,
the CPT symmetry is violated, the
marginal Fermi point splits into topologically-stable Fermi
points.  One can speculate that the latter
happens  for the Standard Model, in particular with the electrically neutral leptons,
the neutrinos
\cite{SplittingPreprint,KlinkhamerJETPL,KlinkhamerNeutrino}. 
The  splitting of Fermi points may also give rise to
a CPT-violating \CSlike~term in the effective gauge field action
\cite{CarollFieldJackiw,AdamKlinkhamerNPB}, as will be discussed later.

Let us first consider this scenario for a marginal
Fermi point describing a \emph{single} pair of relativistic chiral
fermions, that is, one right-handed fermion and one left-handed
fermion. These are Weyl fermions with Hamiltonians $H_{\rm
right}=\vec{\sigma}\cdot{\bf p}$ and $H_{\rm
left}=-\vec{\sigma}\cdot{\bf p}$, where $\vec{\sigma}$ denotes the
triplet of Pauli matrices and natural units are employed with
$c=\hbar=1$. Each of these Hamiltonians has a topologically-stable
Fermi  point ${\bf p}=0$. 
The corresponding inverse Green's functions are given by
\begin{eqnarray} 
G^{-1}_{\rm right}(i\omega,{\bf p})&=&i\omega
-\vec{\sigma}\cdot{\bf p}\;,
\nonumber\\[2mm]
G^{-1}_{\rm left} (i\omega,{\bf p})&=&i\omega +\vec{\sigma}\cdot{\bf p}~.
\label{GreenFWeyl}
\end{eqnarray}
The positions of the Fermi points coincide, ${\bf p}_1={\bf
p}_2=0$, but their topological charges (\ref{TopInvariant}) are
different.  For this simple case, the topological charge equals
the chirality of the fermions, $N_a=C_a$ (i.e., $N=+1$ for the
right-handed fermion and $N=-1$ for the left-handed one).
The total topological charge of the Fermi point ${\bf p}=0$ is
therefore zero.

The splitting of this marginal Fermi point can be described by
the Hamiltonians
$H_{\rm right}=\vec{\sigma}\cdot({\bf
p}-{\bf p}_1)$ and
$H_{\rm left}=-\vec{\sigma}\cdot({\bf p}-{\bf
p}_2)$, with
${\bf p}_1=-{\bf p}_2 \equiv {\bf b}$ from momentum conservation.
The  real vector ${\bf b}$ is assumed to be odd under CPT,
which introduces CPT violation into the physics.
The $4\times 4$ matrix of the combined Green's function has the form
\begin{eqnarray}  
\!\!\!\!\!\!&&\!\!\!\!\!\!
G^{-1}(i\omega,{\bf p}) = \nonumber\\[4mm]
\!\!\!\!\!\!&&\!\!\!\!\!\!
\left(\matrix{i\omega -
\vec{\sigma}\cdot({\bf p}-{\bf b})&0\cr 
0&i\omega+
\vec{\sigma}\cdot({\bf p}+{\bf b})\cr } \right).
\label{ModifiedGreenWeyl}
\end{eqnarray}
    Equation ~(\ref{TopInvariant}) shows that ${\bf p}_1={\bf b}$ is
the Fermi point with topological charge $N=+1$ and
${\bf p}_2=-{\bf b}$
the Fermi point  with topological charge $N=-1$.

Let us now consider the more general situation with both the
electroweak and CPT symmetries broken. The Hamiltonian has then 
an additional mass term,
\begin{eqnarray}
H &=&\left(\matrix{\vec{\sigma}\cdot({\bf p}-{\bf b})&M\cr
       M&-
\vec{\sigma}\cdot({\bf p}+{\bf b})\cr } \right)  \nonumber\\[4mm]
&=&
H_{\rm Dirac} - {\bf I}_{\,2}  \otimes
   \left(\vec{\sigma}\cdot{\bf b}\right) ~.
\label{ModifiedHDirac}
\end{eqnarray}
This Hamiltonian is the typical starting
point for investigations of the effects of
CPT violation in the fermionic sector
(see, e.g., Refs.\ \cite{PerezJHEP,Lehnert} and
references therein). The energy spectrum of Hamiltonian
(\ref{ModifiedHDirac}) is
\begin{eqnarray}
E^2_\pm ({\bf p}) &=& M^2+|{\bf p}|^2+q^2
\nonumber\\[2mm]
&&
\pm \, 2\,q\,\sqrt{M^2+\left({\bf p}\cdot\widehat{{\bf b}}\right)^2}~,
\label{ModifiedEnergyDirac}
\end{eqnarray}
with $\widehat{{\bf b}} \equiv {\bf b}/|{\bf b}|$ 
and $q\equiv |{\bf b}| \geq 0$.

Allowing for a variable parameter $q$, one finds a quantum phase
transition at $q_c=M$ between fully-gapped vacua for $q<M$ and
vacua with two Fermi points for $q>M$. These Fermi points are
given by
\begin{eqnarray} 
{\bf p}_1 &=&+ \widehat{{\bf b}}\;\sqrt{q^2-M^2}  \;,\nonumber\\[2mm]
{\bf p}_2 &=&- \widehat{{\bf b}}\;\sqrt{q^2-M^2}   ~.
\label{FPDiracFermions}
\end{eqnarray}
        Equation ~(\ref{TopInvariant}),
now with a trace over the indices of the $4\times 4$ Dirac
matrices, shows that ${\bf p}_1$ is the Fermi point with
topological charge $N=+1$ and ${\bf p}_2$ the Fermi point with
topological charge $N=-1$ 
[see Fig.~\ref{Fig1} for $\widehat{{\bf b}}=(0,0,1)$]. 
The magnitude of
the splitting of the two Fermi points is given by $2\,\sqrt{q^2-M^2}\,$. At
the quantum phase transition $q_c=M$, the Fermi points with opposite charge
annihilate each other and form a marginal Fermi point 
${\bf p}_0 =0$. The momentum-space topology of this marginal Fermi point is
trivial (topological invariant $N=+1-1=0$).

The full Standard Model contains \emph{eight}  pairs of chiral fermions per
family and a quantum phase transition can be characterized by the
appearance and  splitting of multiple marginal Fermi points. For
systems of cold atoms, an example is provided by
another  spin-triplet $p$--wave state, the so-called 
$\alpha$--phase with orbital antiferromagnetism.  The \BN~Hamiltonian which
qualitatively describes fermionic quasiparticles in the 
$\alpha$--state is given by \cite{VolovikGorkov1985,VollhardtWoelfle}:
\begin{equation}
H= \left(\matrix{ |{\bf p}|^2/ 2m  -q
&   \left( {\bf \Sigma} \cdot {\bf p}\right)        \,c_\perp/\sqrt{3} \cr
    \left( {\bf \Sigma} \cdot {\bf p}\right)^\dagger\,c_\perp/\sqrt{3}  
&-|{\bf p}|^2/ 2m +q \cr } \right),
\label{BogoliubovNambuAlphaH}
\end{equation}
with $|{\bf p}|^2 \equiv p_x^2 +  p_y^2 + p_z^2$
and ${\bf \Sigma} \cdot {\bf p}\equiv 
\sigma_x p_x + \exp(2\pi i/3)\,\sigma_y p_y  +
\exp(-2\pi i/3)\,\sigma_z p_z\,$.

On the BEC side ($q<0$), fermions are again fully-gapped, while on
the BCS side ($q>0$), there are eight Fermi points ${\bf
p}_a$ ($a= 1, \ldots , 8$),  situated at the vertices of a cube in
momentum space \cite{VolovikGorkov1985}. The fermionic excitations
in the vicinity of these points are left- and right-handed Weyl
fermions. In terms of the Cartesian unit vectors ($\widehat{\bf
x}$, $\widehat{\bf y}$, $\widehat{\bf z}$), the four Fermi points
with right-handed Weyl  fermions ($C_a=+1$, for $a= 1, \ldots , 4$) are
given by
\begin{eqnarray} 
{\bf p}_1 &=&p_F\;
(+\widehat{\bf x}+\widehat{\bf y}+\widehat{\bf z})/\sqrt{3}~,
\nonumber\\
{\bf p}_2 &=&p_F\;
(+\widehat{\bf x}-\widehat{\bf y}-\widehat{\bf z})/\sqrt{3}~,
\nonumber\\
{\bf p}_3 &=& p_F\;
(-\widehat{\bf x}-\widehat{\bf y}+\widehat{\bf z})/\sqrt{3}~,
\nonumber\\
{\bf p}_4  &=& p_F\;
(-\widehat{\bf x}+\widehat{\bf y}-\widehat{\bf z})/\sqrt{3}~,
\label{FourFermiPoints}
\end{eqnarray}
while the four Fermi points with 
the left-handed Weyl  fermions ($C_a=-1$, for $a= 5, \ldots , 8$) have
opposite vectors.

Since the quantum phase transition  between the BEC and BCS regimes of
ultracold fermionic atoms and the quantum phase transition for
Dirac fermions with CPT violation are described by 
the same  momentum-space topology, we can expect common properties.
An example of such a common property would be the axial or chiral anomaly.
For quantum anomalies in (3+1)--dimensional systems with Fermi points
and their reduction to (2+1)--dimensional systems, 
see, e.g.,  Refs.~\cite{VolovikBook,StoneRoy} and
references therein. 

One manifestation of the anomaly is
the topological Wess--Zumino--Novikov--Witten  (WZNW) term  in the
effective action.  
The general expression for the WZNW term is represented by the following sum
over Fermi points (see, for example, Eq.~(6a) in
Ref.~\cite{VolovikKonyshev1988}):
\begin{eqnarray}  
\!\!\!\!\!\!\!\!&&\!\!\!\!\!\!\!\! 
S_{\rm WZNW}=(12\pi^2)^{-1}\,\sum_aN_{a}
\nonumber\\
\!\!\!\!\!\!\!\!&&\!\!\!\!\!\!\!\! 
\times\,\int d^3x~d t~d\tau \; {\bf p}_a\cdot
(\partial_\tau{\bf p}_a\times\partial_t{\bf p}_a)\,.
\label{WessZuminoGeneral}
\end{eqnarray} 
Here, $N_a$ is 
the topological charge of the $a$-th Fermi point 
and $\tau \in [0,1]$ is an additional coordinate which parametrizes a
disc, with the usual spacetime at the boundary $\tau=1$.

In the Standard Model, Eq.(\ref{WessZuminoGeneral}) can be seen to 
give rise to an anomalous \CS-like action term in the gauge-field  sector.
Start, for simplicity, from the spectrum of a single electrically charged 
Dirac fermion (charge $e$) and again set $c=\hbar=1$. In the presence of
the vector potential ${\bf A}$ of a $U(1)$ gauge field, the
minimally-coupled version of  Hamiltonian (\ref{ModifiedHDirac}) is 
\begin{eqnarray}  
\!\!\!\!\!\!\!\!&&\!\!\!\!\!\!\!\!
H= \nonumber\\[4mm]
\!\!\!\!\!\!\!\!&&\!\!\!\!\!\!\!\!
\left(\matrix{
        \vec{\sigma}\cdot({\bf p}-e{\bf A} -{\bf b})&M\cr
M&- \vec{\sigma}\cdot({\bf p}-e{\bf A}+{\bf b})\cr }\right)   ~.
\label{ModifiedHDiracGauge}
\end{eqnarray}
The positions of the Fermi points for $q \equiv |{\bf b}| > M$ are
then shifted due to the gauge field,
\begin{eqnarray}
{\bf p}_a &\equiv& {\bf p}_a^{(0)} + e {\bf A} \nonumber\\[2mm]
&=& \pm \widehat{{\bf b}}\;\sqrt{q^2-M^2} + e{\bf A}~,
\label{FPWithGauge}
\end{eqnarray}
with a plus sign for $a=1$ and a minus sign for $a=2$. This result
follows immediately from Eq.~(\ref{FPDiracFermions}) by the
minimal substitution ${\bf p}_a \rightarrow {\bf p}_a - e {\bf A}$, 
consistent with the gauge principle. For \rqfth~and with
different charges $e_a$ at the different Fermi points, one has the
general expression ${\bf p}_a={\bf p}_a^{(0)}+ e_a {\bf A}$.

Next, insert these Fermi points into formula
(\ref{WessZuminoGeneral}) and assume 
the charges to be $\tau$ dependent, 
so that ${\bf p}_a={\bf p}_a^{(0)}+  e_a(\tau)\, {\bf A}$. 
Specifically, we use a parametrization for which the charges
$e_a(\tau)$ are zero at the center of the disc, $e_a(0)=0$, and
equal to the physical charges at the boundary of the disc,
$e_a(1)=e_a$.  From Eq.~(\ref{WessZuminoGeneral}), one then
obtains the general form for the Abelian \CSlike~term
\begin{eqnarray}  
\!\!\!\!\!\!\!\!&&\!\!\!\!\!\!\!\! 
S_{\rm CS-like}=(24\pi^2)^{-1}\,\sum_a N_{a}\,e_a^2\,
\nonumber\\
\!\!\!\!\!\!\!\!&&\!\!\!\!\!\!\!\! 
\times\,
\int d^3x\, d t \;\,  
{\bf p}_a^{(0)} \cdot ({\bf A}\times \partial_t{\bf A})\,.
\label{CSgeneral}
\end{eqnarray} 
This result has the ``relativistic'' form
\begin{equation} 
S_{\rm \,CS-like}= \int d^4x \; k_\mu\,\epsilon^{\mu\nu\rho\sigma}
A_\nu(x)\:\partial_\rho  A_\sigma(x)\,,
\label{CSliketerm}
\end{equation}
with gauge field
$A_\mu(x)$, \LC~symbol $\epsilon^{\mu\nu\rho\sigma}$, and
a purely spacelike ``vector'' $k_\mu$,
\begin{eqnarray}
\!\!\!
k_\mu &=&(0,{\bf k}) \nonumber\\[4mm]
\!\!\!      
      &=&\left(0,\, 
      (24\pi^2)^{-1} \sum_a{\bf p}^{(0)}_a\,e_a^2\,N_a \right)\,.
\label{ManyFermiPointsSpaceLike}
\end{eqnarray}
Note that only gauge invariance has been assumed in the derivation
of Eq.~(\ref{ManyFermiPointsSpaceLike}).
As shown in the Appendix of Ref.~\cite{SplittingPreprint}, the \CS~vector
(\ref{ManyFermiPointsSpaceLike}) can be written in the form
of a momentum-space topological invariant.

Returning to the case of a single Dirac fermion with charge $e$ and
using Eqs.\ (\ref{ManyFermiPointsSpaceLike}) and (\ref{FPDiracFermions}),
one finds that the CPT-violating \CS~parameter
${\bf k}$ can be expressed in terms of the  CPT-violating
parameter ${\bf b}$ of the fermionic sector,
\begin{equation}
{\bf k}={e^2\over 12\pi^2}\;\theta(q - M)\;\widehat{{\bf b}}\;\sqrt{q^2-M^2}  ~.
\label{SpaceLikeKAphase}
\end{equation}
This particular contribution to ${\bf k}$ comes from the splitting of a marginal
Fermi point, which requires $|{\bf b}| \equiv q > M$, as
indicated by the step function on the right-hand side
[$\,\theta(x)=0$ for $x \leq 0$  and $\theta(x)=1$ for $x>0\,$].

In the context  of \rqfth,
the existence of such a nonanalytic contribution to ${\bf k}$
has also been found by Perez-Victoria  \cite{PerezPRL} and Andrianov
et al. \cite{AndrianovJHEP} using
standard regularization methods, but with a prefactor larger by
a factor $3$ and $3/2$, respectively.
The result (\ref{SpaceLikeKAphase}), on the other hand, is determined by
the general topological properties of the Fermi points
\cite{SplittingPreprint} and applies to  nonrelativistic \qfth~as well.
In condensed-matter \qfth, the  result has been obtained without ambiguity,
since the microphysics is  known at all scales and regularization
occurs naturally.

For the ``ferromagnetic'' quantum vacuum of 
Hamiltonian (\ref{ModifiedHDirac}),
the \CS~vector ${\bf k}$ obtained from Eq.(\ref{ManyFermiPointsSpaceLike}) 
by summation over all Fermi points  (\ref{FPDiracFermions}) 
is nonzero and given by Eq.~(\ref{SpaceLikeKAphase}). 
For the ``antiferromagnetic''   $\alpha$--phase vacuum of Hamiltonian
(\ref{BogoliubovNambuAlphaH}), the vector ${\bf k}$ vanishes, 
because $e_a^2=1$ for the fermion charges
$e_a=\pm 1$ and ${\bf p}_1+{\bf p}_2+{\bf p}_3+{\bf p}_4=0$ for
the tetrahedron (\ref{FourFermiPoints}). A similar situation may
occur for the Standard Model: antiferromagnetic splitting of
the Fermi point without induced \CS-like term \cite{SplittingPreprint}. 
The antiferromagnetic splitting may, however, lead to other observable effects
such as neutrino oscillations \cite{KlinkhamerJETPL,KlinkhamerNeutrino}.

In conclusion, one may expect quantum phase transitions in 
systems of ultracold fermionic atoms, 
provided the pairing occurs in the non-$s$-wave
channel. The quantum phase transition separates an 
anomaly-free
fully-gapped vacuum on the BEC side and a gapless superfluid
state on the BCS side, which is characterized by Fermi points and
quantum anomalies. This phenomenon is general and may occur
in many different systems, including the vacuum of the relativistic
quantum field theory relevant to elementary particle physics.


The work of G.E.V. is supported in part
by the Russian  Foundation for Fundamental Research
under grant $\#$02-02-16218
and by the Russian Ministry of Education and
Science, through the Leading Scientific School grant $\#$2338.2003.2 and
the Research Program  ``Cosmion.'' This work is also supported by the
European Science Foundation  COSLAB Program.


\end{document}